\documentclass[12pt]{article}
\usepackage{amssymb,amsmath,epsfig}
\allowdisplaybreaks

\begin{document}
\title{\bf Greybody Factor for a Rotating Bardeen Black Hole}
\author{M. Sharif \thanks{msharif.math@pu.edu.pk} and Qanitah Ama-Tul-Mughani
\thanks{qanitah94@gmail.com}\\
Department of Mathematics, University of the Punjab,\\
Quaid-e-Azam Campus, Lahore-54590, Pakistan.}

\date{}
\maketitle
\begin{abstract}
In this paper, we formulate an analytical expression of greybody
factor in the context of rotating Bardeen black hole which is valid
in low energy and low angular momentum region. Primarily, we analyze
the profile of effective potential which originates the absorption
probability. We then derive two asymptotic solutions by solving the
radial part of Klein-Gordon equation in two different regions,
namely, black hole and far-field horizons. We match these solutions
smoothly to an intermediate regime to extend our results over the
whole radial coordinate. In order to elaborate the significance of
analytical solution, we compute the energy emission rate and
absorption cross-section for the massless scalar field. It is found
that the rotation parameter increases the emission rate of scalar
field particles while the orbital angular momentum minimizes the
emission process.
\end{abstract}

\section{Introduction}

Black holes (BHs) are the most mysterious and abstruse astronomical
objects as they have singularities covered by the event horizons. In
general relativity, singularity is a region where all laws of
physics break down and gravitational pull diverges in this region of
spacetime. To avoid these undefined regions, a set of solutions
dubbed as ``regular BHs" have played an essential role to overcome
this obstacle which do not have any singularity even at the origin.
In literature, the first ever regular spherically symmetric BH
solution was obtained by Bardeen \cite{2}, known as Bardeen BH. As
the Bardeen BH was not a vacuum solution, therefore a special form
of the energy-momentum tensor was introduced to attain the model
satisfying the weak energy bounds. After that, characteristics of
such regular BH solutions have extensively been carried out in
general relativity \cite{2'}-\cite{9}.

Newman and Janis \cite{10} applied a complex coordinate
transformation on Schwarzschild solution to obtain a regular as well
as rotating BH solution without using the field equations.
Ay$\acute{o}$n-Beato and Garc$\acute{i}$a \cite{3} reinterpreted the
Bardeen model as the magnetic solution of field equations coupled
with nonlinear electrodynamics. Drake and Szekeres \cite{11}
generated the Kerr-Newman metric, representing a rotating charged
BH, by applying Newman-Janis scheme on the
Reissner-Nordstr$\ddot{o}$m metric. Bambi and Modesto \cite{12} used
this algorithm to the regular Hayward and Bardeen BH metrics to
construct a family of rotating BH spacetimes.

Classically, it is accepted that BH absorbs everything from its
surrounding due to strong gravitational effects and does not emit
any type of radiations as well as particles. However, it has been
shown that BHs can create and discharge thermal radiations by taking
into account the quantum mechanical effects. In the gravitational
context, these thermal emissions named as Hawking radiation \cite{1}
gradually lead to decrease in the BH mass and finally to its
eventual evaporation. It is believed that any primordial BH having
mass less than $10^{15}$g would have disappeared by now. Black holes
as thermal objects associate entropy and Hawking temperature which
vary for different class of BHs \cite{13}-\cite{15}. The emission
process at the horizon of BH, in terms of frequency, is given as
\cite {1}
\begin{equation}\nonumber
\gamma(w)=\left(\frac{d^{3}k}{e^{\frac{w}{T_{H}}}(2\pi)^3}\right),
\end{equation}
where $T_{H}$ is the Hawking temperature. This expression can be
modified up to any dimensions and valid for massless as well as
massive particles. Consequently, the radiation spectrum at the
horizon is exactly equal to the black body spectrum which gives rise
to loss of information paradox. The inevitable reality is that the
spacetime around a BH is nontrivial which alters the spectra of
Hawking radiation by allowing some of the radiations to cross the
barrier and rest of them reflect back to BH.

The greybody factor (rate of absorption probability) is defined as
the probability for an incoming wave from infinity to be absorbed by
the BH which is directly related to absorption cross-section \cite
{15'}-\cite{18}. The relation between emission rate and greybody
factor can be expressed as
\begin{equation}\nonumber
\gamma(w)=\left(\frac{|\tilde{\emph{A}}_{l,m}|^{2}
d^{3}k}{e^{\frac{w}{T_{H}}}(2\pi)^3}\right),
\end{equation}
where $|\tilde{\emph{A}}_{l,m}|^{2}$ depicts the greybody factor
which is also a frequency dependent quantity. Creek et al.
\cite{40,41} studied the scalar emission rate for rotating BH and
derived both numerical as well as analytical solutions for the
greybody factor. Boonserm et al. \cite{41'} computed some rigorous
bounds on the greybody factors associated with scalar field
excitations for the Myers-Perry BH. Jorge et al. \cite{41''}
explored the greybody factors in low frequency regime for higher
dimensional rotating BHs with the cosmological constant being zero,
positive or negative. Toshmatov et al. \cite{42} investigated the
absorption probability for regular BH spacetimes and observed that
charge parameter decreases the transmission rate of the incident
wave. Ahmed and Saifullah \cite{19} computed the greybody factor for
uncharged scalar particles in the background of cylindrically
symmetric spacetime and found an analytical solution in the form of
hypergeometric function. Dey and Chakrabarti \cite{20} evaluated the
quasinormal modes as well as absorption probability for the
Bardeen-de Sitter BH due to gravitational perturbation. Recently,
Hyun et al. \cite{20'} found an analytical expression of greybody
factor for the brane scalar field in five-dimensional rotating BH by
using spheroidal joining factor.

In this paper, we derive an analytical solution of absorption
probability for rotating Bardeen BH. The paper is organized as
follows. In section \textbf{2}, we evaluate effective potential
through the decoupled set of radial and angular equations resulting
from the Klein-Gordon equation. Section \textbf{3} leads to two
asymptotic solutions by solving the radial equation of motion at two
different horizons. We also calculate the energy emission rate and
absorption cross-section for the massless scalar field. The last
section concludes our results.

\section{Klein-Gordon Equation and Effective Potential}

The axially symmetric spacetime for rotating Bardeen BH in
Boyer-Lindquist coordinates is given as
\begin{equation}\label{1}
ds^{2}=-F(r,\theta)dt^{2}+\frac{1}{G(r,\theta)}dr^{2}+\Sigma(r,\theta)
d\theta^2+H(r,\theta)d\phi^2-2K(r,\theta)dtd\phi,
\end{equation}
where
\begin{eqnarray}\nonumber
F(r,\theta)&=&\frac{\Pi(r)-a_{0}^{2}\sin^{2}\theta}{\Sigma(r,\theta)},
\quad G(r,\theta)=\frac{\Pi(r)}{\Sigma(r,\theta)}, \quad
\Sigma(r,\theta)=r^{2}+a_{0}^{2}\cos^{2}\theta,\\
\nonumber
H(r,\theta)&=&\frac{\sin^{2}\theta\left((r^{2}+a_{0}^{2})^{2}
-\Pi(r)a_{0}^{2}\sin^{2}\theta\right)}{\Sigma(r,\theta)},\quad
\Pi(r)=r^{2}+a_{0}^{2}-2rM(r),\\ \nonumber
K(r,\theta)&=&\frac{a_{0}\sin^{2}\theta(r^{2}+a_{0}^{2}-\Pi(r))}{\Sigma(r,\theta)},
\end{eqnarray}
$a_{0}$ represents the rotation parameter while the mass function is
defined as
\begin{equation}\nonumber
M(r)=M_{0}\left(\frac{r^3}{(r^{2}+r_{0}^{2})^{\frac{3}{2}}}\right).
\end{equation}
Here, $M_{0}$ corresponds to the mass of BH and $r_{0}$ as length
parameter, measures the deviation of considered class of BH from the
Kerr geometry \cite{12}. Moreover, it is noted that curvature
invariants, i.e., Ricci and Kretschmann scalars are regular (bounded
and non-singular) everywhere even at the origin, hence we can name
the line element (\ref{1}) as rotating Bardeen regular BH. Through
computing $R_{\mu\nu}$ and $T_{\mu\nu}$, Azreg-Ainou \cite{20''}
found that the spacetime (\ref{1}) satisfies the field equations.
The event horizons are evaluated by the condition
\begin{equation}\label{1'}
\Pi(r)=0=r^{2}-2rM(r)+a_{0}^{2}.
\end{equation}

In order to calculate the analytic expression of greybody factor, we
first derive the equation of motion to examine the propagation of
scalar field. We presume that particles are only minimally coupled
to gravity and do not involve in any other interaction. In this
scenario, the equation of motion takes the form
\begin{equation}\label{2}
\nabla_{\mu}\nabla^{\mu}\Psi=\partial_{\mu}[\sqrt{-g}g^{\mu\nu}
\partial_{\nu}\Psi(t,r,\theta,\phi)]=0,
\end{equation}
which, through Eq.(\ref{1}) and
$\sqrt{-g}=-\frac{\Sigma(K^2+FH)}{G}$, turns out to be
\begin{eqnarray}\nonumber
&\sqrt{-g}\left(\frac{-H}{K^2+FH}\right)\partial_{tt}\Psi
+2\sqrt{-g}\left(\frac{-K}{K^2+FH}\right)\partial_{t}\partial_{\phi}\Psi
+\sqrt{-g}\left(\frac{F}{K^2+FH}\right)\partial_{\phi\phi}\Psi&\\
\label{3}&
+(\sqrt{-g}G\partial_{r}\Psi)_{,r}+(\sqrt{-g}\frac{1}{\Sigma}
\partial_{\theta}\Psi)_{,\theta}=0.&
\end{eqnarray}
For separation of variables, the field factorization
\begin{equation}\nonumber
\Psi(t,r,\theta,\phi)=\exp(-\iota wt)\exp(\iota
m\phi)R_{wlm}(r)S^{m}_{l}(\theta,a_{0}w),
\end{equation}
where $S^{m}_{l}(\theta,a_{0}w)$ are angular spheroidal functions
\cite{24, 24a}, leads to the following decoupled set of radial and
angular equations
\begin{eqnarray}\nonumber
&\frac{\partial}{\partial r}(\Pi\frac{\partial R_{wlm}}{\partial r})
+\left[\frac{1}{\Pi}((w^2(r^2+a_{0}^{2})^2-a_{0}^2)-a_{0}^2m^{2}-4a_{0}rM)
-\lambda_{l}^{m}\right]R_{wlm}=0,&\\\label{5} \\ \label{6}
&\frac{1}{\sin\theta}\frac{\partial
}{\partial\theta}\left(\sin\theta \frac{\partial
S^{m}_{l}}{\partial\theta}\right)+(\frac{m^{2}}{\sin^{2}\theta}
-w^{2}a_{0}^{2}\sin^{2}\theta+\lambda_{l}^{m})S^{m}_{l}(\theta,a_{0}w)=0.&
\end{eqnarray}
Here $\lambda_{l}^{m}$ is a separation constant which establishes a
connection between decoupled equations.

In general, the separation constant cannot be expressed in a closed
form. However, its analytic expression can be obtained as a power
series in terms of parameter $a_{0}w$ \cite{22, 22a} given as
\begin{equation}\label{7}
\lambda_{l}^{m}=\sum_{n=0}^{\infty}\textit{f}_{~n}^{~lm}(a_{0}w)^{n}.
\end{equation}
For the purpose of simplicity, we truncate the series and only keep
the terms up to third order which is expressed as follows
\begin{equation}\label{8}
\lambda_{l}^{m}=l(l+1)+\frac{-2m^{2}+2l(l+1)-1}{(2l-1)(2l+3)}
(a_{0}w)^{2}+O((a_{0}w)^{4}),
\end{equation}
with $\textit{f}_{~1}^{~lm}=\textit{f}_{~3}^{~lm}=0$. Here, the
parameter $l$ denotes the orbital angular momentum with non-negative
integeral values satisfying $l\geq |m|$ and $\frac{l-|m|}{2}\in
\{0,\mathbb{Z}^{+}\}$. Now, we are in a position to solve the radial
equation (\ref{5}) analytically by using the above mentioned power
series expression. The obtained solution will lead us to determine
the greybody factor for a massless scalar field. Before computing
greybody factor, we analyze the profile of effective potential which
is responsible to originate it. Introducing a new radial mode
transformation
\begin{equation}\label{9}
R_{wlm}(r)=\frac{U_{wlm}(r)}{\sqrt{r^2+a_{0}^{2}}},
\end{equation}
and tortoise coordinate $x_{*}$
\begin{equation}\label{10}
\frac{dx_{*}}{dr}=\frac{r^{2}+a_{0}^2}{\Pi(r)},
\end{equation}
such that
\begin{equation}\label{11}
\frac{d}{dx_{*}}=\frac{\Pi(r)}{r^{2}+a_{0}^2}\frac{d}{dr}, \quad
\frac{d^2}{dx_{*}^{2}}=\left(\frac{\Pi(r)}{r^{2}+a_{0}^2}\right)^2
\frac{d^2}{dr^2}+\left(\frac{\Pi(r)}{r^{2}+a_{0}^2}\right)\frac{d}{dr}
\left(\frac{\Pi(r)}{r^{2}+a_{0}^2}\right)\frac{d}{dr}.
\end{equation}
It is observed that as $r$ approaches to event horizon, i.e.,
$r\rightarrow r_{h}$ the tortoise coordinate approaches to $-\infty$
whereas, for $r\rightarrow \infty$ we get $x_{*}\rightarrow \infty$.
Thus, the Regge-Wheeler equation (\ref{5}) is confined only for
zones located outside the horizon of BH whereas tortoise coordinate
maps it onto the entire real line. In this context, Eq.(\ref{5})
takes the form of standard Schr$\ddot{o}$dinger equation
\begin{equation}\label{12}
(\frac{d^2}{dx_{*}^2}-V_{eff})U_{wlm}(r)=0,
\end{equation}
where
\begin{eqnarray}\nonumber
V_{eff}&=&\{\frac{d}{dr}\left[\frac{r\Pi(r)}{(r^{2}+a_{0}^2)^{\frac{3}{2}}}\right]
-\frac{1}{\Pi}[w^2(r^2+a_{0}^{2})^2-a_{0}^2m^{2}-4a_{0}rM(r)]\\
\label{13} &+& l(l+1)
-\frac{2m^{2}-2l(l+1)+1}{(2l-1)(2l+3)}(a_{0}w)^{2}\}\frac{\Pi(r)}{(r^{2}+a_{0}^2)^2}.
\end{eqnarray}
\begin{figure}
  \includegraphics[width=3 in, height=2.5 in]{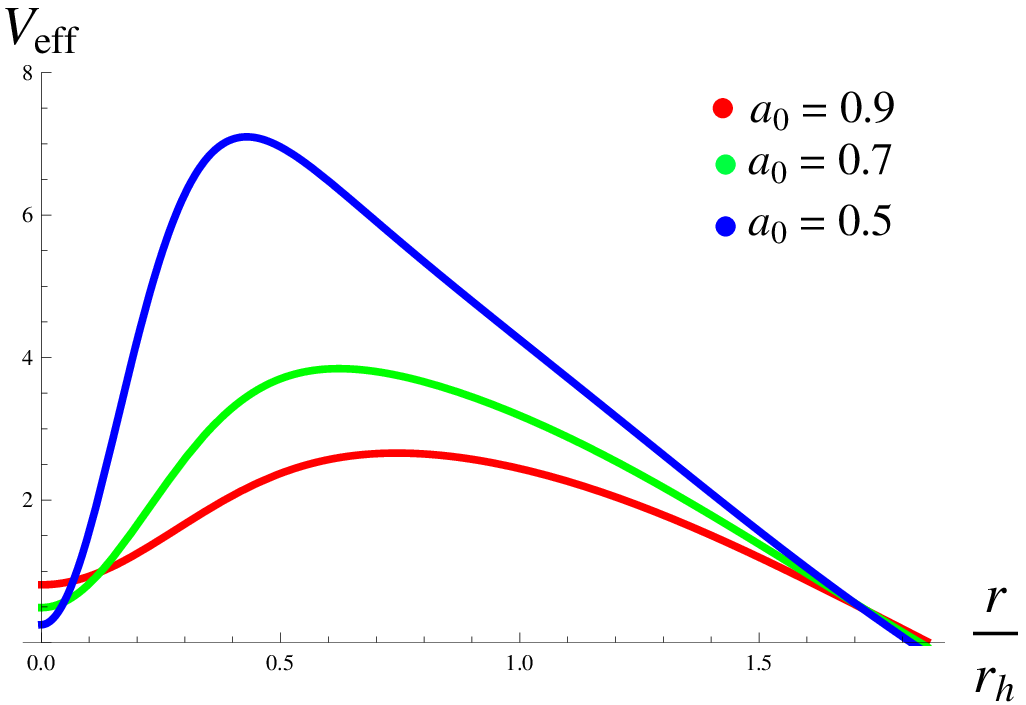}
    \includegraphics[width=3 in, height=2.5 in]{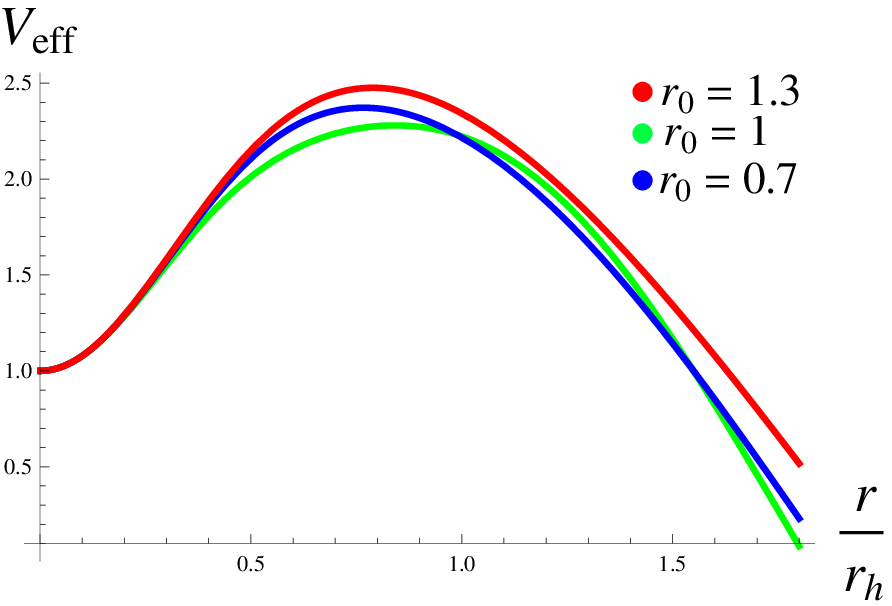}\\
 \caption{Effective potential for massless scalar field corresponding
to $r_{0}=1$ (left plot) and  $a_{0}=1$ (right plot) with $m=1$,
$l=2$ and $w=0.1$.}\label{1}
\end{figure}
\begin{figure}
  \includegraphics[width=3 in, height=2.5 in]{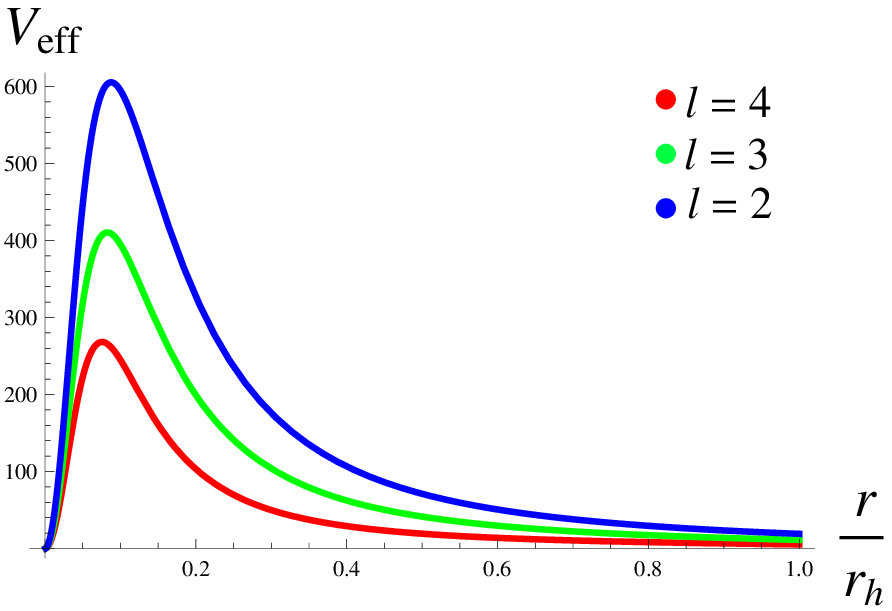}
    \includegraphics[width=3 in, height=2.5 in]{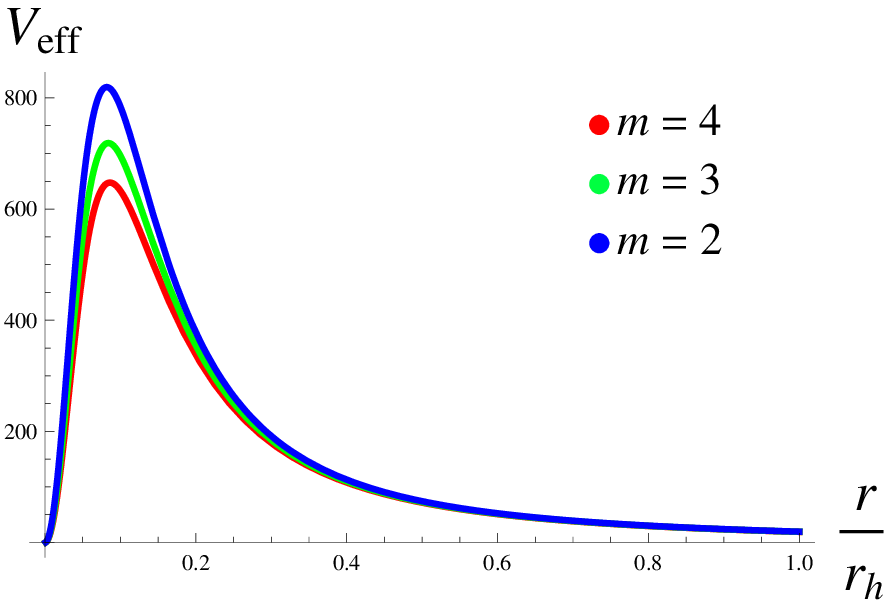}\\
 \caption{Effective potential for massless scalar field corresponding
to $m=1$ (left plot) and  $l=4$ (right plot) with $a_{0}=w=0.1$ and
$r_{0}=1$.}\label{2}
\end{figure}

It is noted that the effective potential turns out to be zero at
event horizons due to the vanishing of metric coefficient
$G(r,\theta)$. For graphical analysis, we plot the effective
potential versus $\frac{r}{r_{h}}$ to examine the impact of
different parameters on gravitational barrier as shown in Figures
\textbf{1-2}. Firstly, we set the parameters $r_{0}=1=m$, $ l=2$,
$w=0.1$ and sketch the graphs for different choices of the rotation
parameter. It is found that the height of potential barrier
decreases with the increase in $a_{0}$  which causes an enhancement
in the emission of scalar fields (left plot of Figure \textbf{1}).
As a result, the greybody factor increases significantly
corresponding to larger values of rotation parameter. In the right
plot of Figure \textbf{1}, we display the effect of length parameter
on the profile of effective potential. We observe that the
gravitational potential is positive, finite and has a direct
relation with $r_{0}$, i.e., increase in the values of $r_{0}$ leads
to increase in the height of barrier which subsequently decreases
the emission of scalar particles. The dependence of angular momentum
numbers is also demonstrated in Figure \textbf{2} which shows that
higher spikes of the effective potential are attained for larger
values of $l$ and $m$ which ultimately minimize the emission process
and greybody factor.

\section{Greybody factor}

In this section, we obtain an analytical solution of the greybody
factor in low energy regime by using an approximation technique. We
solve Eq.(\ref{5}) for two asymptotic regions separately, i.e.,
close to BH horizon and far-away from it. These two solutions will
be extended and matched smoothly in an intermediate region to get an
analytical expression for the whole radial regime.

\subsection{Analytical solution}

For the near horizon regime $r\sim r_{h}$, we apply the
transformation
\begin{equation}\label{14}
r\rightarrow h=\frac{\Pi(r)}{r^2+a_{0}^{2}},
\end{equation}
such that
\begin{equation}\label{15}
\frac{dh}{dr}=\frac{(1-h)D(r_{h})}{r_{h}(r_{h}^{2}+a_{0}^2)(r_{h}^{2}+r_{0}^2)},
\end{equation}
where
\begin{equation}\label{16}
D(r_{h})=r_{h}^4-a_{0}^{2}r_{h}^2-2r_{0}^{2}r_{h}^
2-4r_{0}^{2}a_{0}^2.
\end{equation}
Using the above expressions in the radial equation of motion, we
obtain
\begin{eqnarray}\label{17}
&h(1-h)\frac{d^{2}R_{wlm}}{dh^{2}}+(1-C_{*}h)\frac{dR_{wlm}}{dh}
+[\frac{\chi^{*}}{D^{2}(r_{h})(1-h)h}-\frac{\lambda_{h}^{*}}{D^{2}(r_{h})(1-h)}]=0,&
\end{eqnarray}
where
\begin{eqnarray}\label{18}
C_{*}&=&-\frac{(r_{h}^{2}+4r_{0}^{2})(r_{h}^{2}+a_{0}^{2})}{D(r_{h})},\\
\label{19}
\chi^{*}&=&r_{h}^{2}(r_{h}^{2}+r_{0}^2)^{2}[w^2(r^2+a_{0}^{2})^2
-a_{0}^{2}m^{2}-4a_{0}rM(r)],
\end{eqnarray}
and
\begin{eqnarray}\label{20}
\lambda_{h}^{*}&=&r_{h}^{2}(r_{h}^{2}+a_{0}^2)(r_{h}^{2}+r_{0}^2)^{2}[l(l+1)
-\frac{2m^{2}-2l(l+1)+1}{(2l-1)(2l+3)}(a_{0}w)^{2}].
\end{eqnarray}

In order to get a hypergeometric equation, we use the field
redefinition
\begin{equation}\label{21}
R_{wlm}(h)=h^{\mu_{1}}(1-h)^{\nu_{1}}\tilde{F}(h),
\end{equation}
which reduces Eq.(\ref{17}) to
\begin{eqnarray}\nonumber
&h(1-h)\frac{d^{2}\tilde{F}}{dh^{2}}+[1+2\mu_{1}-(2\mu_{1}+2\nu_{1}
+C_{*})h]\frac{d\tilde{F}}{dh} +(\frac{\mu_{1}^2}{h}-\mu_{1}^2&\\
\nonumber&+\mu_{1}
-2\mu_{1}\nu_{1}-\nu_{1}^2+\frac{\nu_{1}^2}{1-h}-\frac{2\nu_{1}}{1-h}
+\nu_{1}-\mu_{1}C_{*}+\frac{\nu_{1}C_{*}}{1-h}&\\
\label{24}&-\nu_{1}C_{*}
+\frac{\chi^{*}}{D^{2}h}+\frac{\chi^{*}}{D^{2}(1-h)}
-\frac{\lambda_{h}^{*}}{D^{2}(1-h)})\tilde{F}=0.&
\end{eqnarray}
Here
\begin{eqnarray}\label{25}
\tilde{a}=\mu_{1}+\nu_{1}+C_{*}-1, \quad \tilde{b}=\mu_{1}+\nu_{1},
\quad \tilde{c}=1+2\mu_{1}.
\end{eqnarray}
The power coefficients $\mu_{1}$ and $\nu_{1}$ can be calculated by
second order algebraic equations (following the condition that
coefficient of $\tilde{F}(h)$ must be $-\tilde{a}\tilde{b}$),
namely,
\begin{eqnarray}\label{26}
\mu_{1}^{2}+\frac{\chi^{*}}{D^{2}}&=&0,\\\label{27}
\nu_{1}^{2}+\nu_{1}(C_{*}-2)+\frac{\chi^{*}}{D^{2}}
-\frac{\lambda_{h}^{*}}{D^{2}}&=&0.
\end{eqnarray}
The radial equation combined with Eq.(\ref{25}) and constraints
(\ref{26}, \ref{27}) leads to
\begin{equation}\label{28}
h(1-h)\frac{d^{2}\tilde{F}}{dh^2}+[\tilde{c}-(1+\tilde{a}+\tilde{b})h]
\frac{d\tilde{F}}{dh}-\tilde{a}\tilde{b}\tilde{F}(h)=0.
\end{equation}
Thus, the general solution of Eq.(\ref{24}) in terms of
hypergeometric function can be expressed as
\begin{eqnarray}\nonumber
(R_{wlm})_{NH}(h)&=&A_{1}h^{\mu_{1}}(1-h)^{\nu_{1}}\tilde{F}(\tilde{a},\tilde{b},\tilde{c};h)\\
\label{29}
&+&A_{2}h^{-\mu_{1}}(1-h)^{\nu_{1}}\tilde{F}(\tilde{a}-\tilde{c}+1,\tilde{b}-\tilde{c}+1,2-\tilde{c};h),
\end{eqnarray}
where $A_{1}$ and $A_{2}$ are arbitrary constants with
\begin{eqnarray}\label{29'}
\mu_{1}^{\pm}&=&\pm\iota\frac{\chi^{*}}{D(r_{h})},\\ \label{30}
\nu_{1}^{\pm}&=&\frac{1}{2}[(2-C_{*})\pm\sqrt{(2-C_{*})^2-4(\frac{\chi^{*}}{D^2}
-\frac{\lambda_{h}^{*}}{D^{2}})}].
\end{eqnarray}

Applying the boundary constraint that no outgoing mode exists near
the horizon of BH, we can take either $A_{1}=0$ or $A_{2}=0$
depending upon the signature of $\mu_{1}$. As these two constants
become equivalent for both choices of $\mu_{1}$, so we opt
$\mu_{1}=\mu_{1}^{-}$ and choose $A_{2}=0$. Similarly, the sign of
$\nu_{1}$ can be decided by using the convergence property of
hypergeometric function which is satisfied for
$\nu_{1}=\nu_{1}^{-}$. Hence the final expression of near horizon
solution is given as
\begin{equation}\label{30'}
(R_{wlm})_{NH}(h)=A_{1}h^{\mu_{1}}(1-h)^{\nu_{1}}\tilde{F}(\tilde{a},\tilde{b},\tilde{c};h).
\end{equation}
Our next goal is to evaluate the solution of radial equation in the
far-field regime. Using the assumption $r>>r_{h}$ and keeping the
leading factor in the expansion of $\frac{1}{r}$, Eq.(\ref{5}) takes
the form
\begin{equation}\label{31'}
\frac{d^{2}\tilde{R}_{wlm}}{dr^2}+\frac{1}{r}\frac{d\tilde{R}_{wlm}}{dr}
+\{w^2-\frac{1}{r^2}\frac{[l(l+1)+2(a_{0}w)^2(l(l+1)-1)]}{(2l-1)(2l+3)}
+\frac{1}{4}\}\tilde{R}_{wlm}=0,
\end{equation}
which is known as Bessel equation with
$R_{wlm}(r)=\frac{1}{\sqrt{r}}\tilde{R}_{wlm}$. Thus, the solution
in the far-field region is given as
\begin{equation}\label{31}
(R_{wlm})_{FF}=\frac{\tilde{B}_{1}}{\sqrt{r}}J_{v}(wr)+\frac{\tilde{B}_{2}}{\sqrt{r}}Y_{v}(wr),
\end{equation}
where $B_{1, 2}$ are integration constants, $J_{v}(wr)$ and
$Y_{v}(wr)$ are the Bessel functions with
$v=\sqrt{l(l+1)+2(a_{0}w)^2\frac{(l(l+1)-1)}{(2l-1)(2l+3)}+\frac{1}{4}}$.

\section{Matching to an Intermediate Regime}

In order to obtain an analytical solution which remains valid for
complete range of $r$, we have to match the obtained solutions
smoothly in the intermediate region. For this purpose, we first
stretch the near horizon solution by exchanging the argument of
hypergeometric function from $h$ to $1-h$ as
\begin{eqnarray}\nonumber
(R_{wlm})_{NH}(h)&=&A_{1}h^{\mu_{1}}(1-h)^{\nu_{1}}\{\frac{\Gamma(\tilde{c})
\Gamma(\tilde{c}-\tilde{a}-\tilde{b})}{\Gamma(\tilde{c}-\tilde{a})\Gamma(\tilde{c}-\tilde{b})}
\tilde{F}(\tilde{a},\tilde{b},\tilde{c};1-h)\\ \nonumber &+&
(1-h)^{\tilde{c}-\tilde{a}-\tilde{b}}\frac{\Gamma(\tilde{c})
\Gamma(\tilde{a}+\tilde{b}-\tilde{c})}{\Gamma(\tilde{a})\Gamma(\tilde{b})}\\\label{32}&\times&
\tilde{F}(\tilde{c}-\tilde{a},-\tilde{b}+\tilde{c},\tilde{c}-\tilde{b}-\tilde{a}+1;1-h)\}.
\end{eqnarray}
Using horizon equation, the function $h(r)$ can be expressed as
\begin{equation}\label{33}
h(r)=1-\frac{r}{r_{h}}\left[\frac{1+a_{*}^{2}}{(\frac{r}{r_{h}})^2+a_{*}^2}\right],
\end{equation}
with $a_{*}= \frac{a_{0}}{r_{h}}$. When $r>>r_{h}$, the term
$(\frac{r}{r_{h}})^2$, appearing in the second factor, becomes
dominant which leads the whole expression towards unity. In the
limiting value $h\rightarrow 1$, we have
\begin{equation}\label{34}
(1-h)^{\nu_{1}}\simeq(\frac{r_{h}}{r})^{\nu_{1}},
\end{equation}
and
\begin{equation}\label{35}
(1-h)^{\nu_{1}+\tilde{c}-\tilde{a}-\tilde{b}}\simeq(\frac{r_{h}}{r})^{-\nu_{1}+2-C_{*}}.
\end{equation}
Thus, the near horizon solution in an intermediate region becomes
\begin{equation}\label{37}
(R_{wlm})_{NH}(r)=A^{*}_{1}r^{-\nu_{1}}+A^{*}_{2}{r}^{\nu_{1}-2+C_{*}},
\end{equation}
with
\begin{eqnarray}\label{38}
A^{*}_{1}&=&A_{1}[(1+a_{*}^2)r_{h}]^{\nu_{1}}\frac{\Gamma(\tilde{c})
\Gamma(\tilde{c}-\tilde{a}-\tilde{b})}{\Gamma(\tilde{c}-\tilde{a})\Gamma(\tilde{c}-\tilde{b})},\\\label{39}
A^{*}_{2}&=&A_{1}[(1+a_{*}^2)r_{h}]^{(-\nu_{1}-C_{*}+2)}\frac{\Gamma(\tilde{c})
\Gamma(\tilde{a}+\tilde{b}-\tilde{c})}{\Gamma(\tilde{a})\Gamma(\tilde{b})}.
\end{eqnarray}
Setting the limit $wr\rightarrow 0$ in Eq.(\ref{31}) to shift the
far-field solution towards the smaller values of $r$, we obtain
\begin{eqnarray}\nonumber
(R_{wlm})_{FF}&\simeq&\frac{B_{1}(\frac{wr}{2})^{\sqrt{\lambda_{l}^{m}
+\frac{1}{4}-w^2a_{0}^2}}}
{\sqrt{r}\Gamma(\sqrt{\lambda_{l}^{m}+\frac{1}{4}-w^2a_{0}^2}+1)}\\
\label{40}
&-&\frac{B_{2}\Gamma(\sqrt{\lambda_{l}^{m}+\frac{1}{4}-w^2a_{0}^2})
(\frac{wr}{2})^{-\sqrt{\lambda_{l}^{m}+\frac{1}{4}-w^2a_{0}^2}}}
{\sqrt{r}\pi}.
\end{eqnarray}

We observe that power coefficients of the radial component in both
solutions are different from each other which restrain the exact
matching. To resolve this problem, the power-law expressions are
expanded in the limit $(\frac{a_{0}}{r_{h}})^2<<1$ and
$(wr_{h})^2<<1$. These approximations restrict the accuracy of our
solutions in the low rotation and low energy region. In this
expansion, we also ignore the second order term $(a_{0}w)^{2}$ to
achieve the smooth matching. Employing these restrictions,
Eq.(\ref{37}) takes the form
\begin{eqnarray}\label{41}
-\nu_{1}&=&l+O(w_{*}^{2},a_{*}^{2},w_{*}a_{*}),\\\label{42}
\nu_{1}-2+C_{*}&=&-(l+1)+O(w_{*}^{2},a_{*}^{2},w_{*}a_{*}).
\end{eqnarray}
Similarly, the power coefficient in Eq.(\ref{40}) reduces to
\begin{equation}\label{43}
\sqrt{\lambda_{l}^{m}+\frac{1}{4}-w^2a_{0}^2}=(l+\frac{1}{2})+O(w_{*}^{2}a_{*}^{2}).
\end{equation}
It is worth mentioning here that all aforementioned approximations
are not used in the argument of gamma function to increase the
efficiency of our results. One can easily check that both asymptotic
solutions have the same power coefficients, i.e., $r^{l}$ and
$r^{-(l+1)}$. Therefore, we can determine the integration constants
by comparing the corresponding coefficients of Eqs.(\ref{37}) and
(\ref{40}). Thus, the matching of these two solutions leads to
\begin{eqnarray}\nonumber
\tilde{B}&=&\frac{\tilde{B}_{1}}{\tilde{B}_{2}}=\frac{1}{\pi}(\frac{2}{w(1+a_{*}^{2})r_{h}})^{2l+1}
\sqrt{\lambda_{l}^{m}+\frac{1}{4}-w^2a_{0}^2}\\ \label{44}&\times&
\frac{\Gamma^{2}(\sqrt{\lambda_{l}^{m}+\frac{1}{4}-w^2a_{0}^2})
\Gamma(\tilde{a})\Gamma(\tilde{b})\Gamma(\tilde{c}-\tilde{a}-\tilde{b})}{\Gamma(\tilde{c}-\tilde{a})
\Gamma(\tilde{c}-\tilde{b})\Gamma(\tilde{a}+\tilde{b}-\tilde{c})},
\end{eqnarray}
which ensures an analytical smooth solution of the radial equation
for all choices of $r$, valid in low angular momentum and low energy
region.

Finally, to compute the greybody factor, we stretch Eq.(\ref{31}) to
$r\rightarrow \infty$ which leads to
\begin{eqnarray}\nonumber
(R_{wlm})_{FF}(r)&\simeq&\frac{1}{\sqrt{2\pi
w}}\left[\frac{\tilde{B}_{1}
+\iota \tilde{B}_{2}}{r}\exp{-\iota(wr-\frac{\pi}{2}v-\frac{\pi}{4})}\right.\\
\label{45}&+&\left.\frac{\tilde{B}_{1}-\iota
\tilde{B}_{2}}{r}\exp{\iota(wr-\frac{\pi}{2}v
-\frac{\pi}{4})}\right],\\\label{46}&=&\tilde{A}_{in}^{(\infty)}\frac{\exp^{-\iota
wr}}{r}+\tilde{A}_{out}^{(\infty)} \frac{\exp^{\iota wr}}{r}.
\end{eqnarray}
The effects of rotation parameter become negligible at a large
distance from BH. This reduces our massless scalar field solution to
a spherical wave \cite{25}-\cite{25b} allowing to calculate the
greybody factor as
\begin{eqnarray}\label{47}
|\tilde{\emph{A}}_{l,m}|^{2} &=&
1-\left|\frac{\tilde{A}_{out}^{(\infty)}}{\tilde{A}_{in}^{(\infty)}}
\right|^{2}=1-\left|\frac{\tilde{B}_{1}-\iota
\tilde{B}_{2}}{\tilde{B}_{1}+\iota \tilde{B}_{2}}\right|^{2}, \\
\label{48} &=&
1-\left|\frac{\tilde{B}-\iota}{\tilde{B}+\iota}\right|^{2}=\frac{2\iota(\tilde{B}^{*}-\tilde{B})}
{\tilde{B}\tilde{B}^{*}+\iota(\tilde{B}^{*}-\tilde{B})+1}.
\end{eqnarray}
The above expression, together with  Eq.(\ref{44}), is the main
result for the greybody factor specifying the emission of massless
scalar field from a rotating regular BH.

Since any traveling wave passing the far-field horizon will face the
potential barrier as a hindrance, therefore some of its part will be
transmitted towards the horizon of BH and some is reflected back to
the far-field regime. Basically, it is the relative relation between
the effective potential and frequency which decides either to
reflect back the wave or move forward. When the frequency of the
wave is larger than the effective barrier, it can cross the barrier
and will not be reflected back. In the reverse case, when the height
of potential is larger as compared to the frequency, most of the
part will be reflected towards the far-field region and some of its
portion may cross the barrier through the tunneling effect. In this
scenario, the greybody factor shows a negative trend.

To study the characteristics of greybody factor in detail, we sketch
our main expressions, given by Eqs.(\ref{44}) and (\ref{48}), for
scalar particle $(l, m)$ and spacetime $(a_{0}, r_{0})$ variables.
It is noted that the greybody factor remains positive throughout the
considered domain of dimensionless parameter $wr_{h}$. In Figure
\textbf{3}, we fix the spacetime variables and plot the graphs for
discrete choices of angular momentum numbers. It is found that the
increase in the values of $m$ leads to larger ranges of the greybody
factor (left plot) whereas the reverse effects are observed for
orbital angular momentum (right plot). One can easily observe that
the partial wave with $l=1$ dominates in the low energy region while
all higher modes strongly minimize the absorption probability. The
impact of $r_{0}$ and $a_{0}$ on the greybody factor are shown in
Figure \textbf{4}. We note that increase in the values of rotation
as well as length parameter yields the higher values of absorption
probability.
\begin{figure}
  \includegraphics[width=3 in, height=2.5 in]{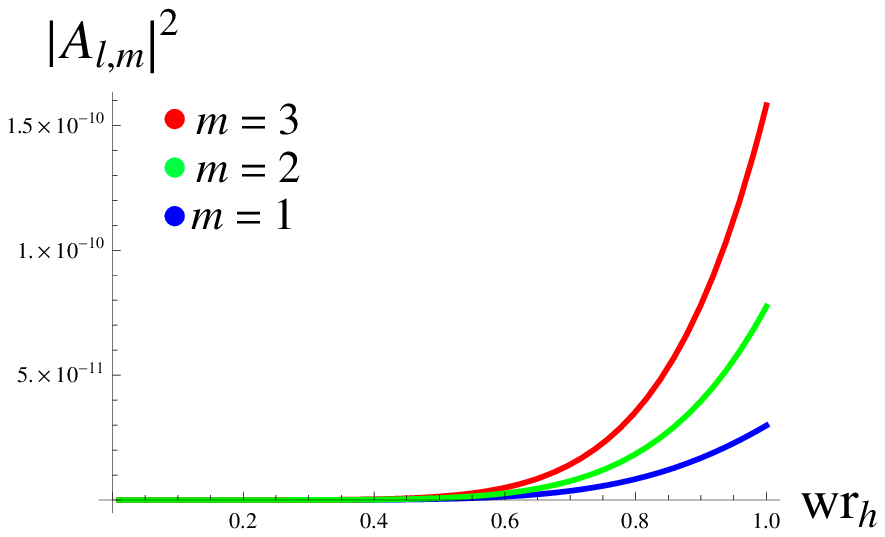}
    \includegraphics[width=3 in, height=2.5 in]{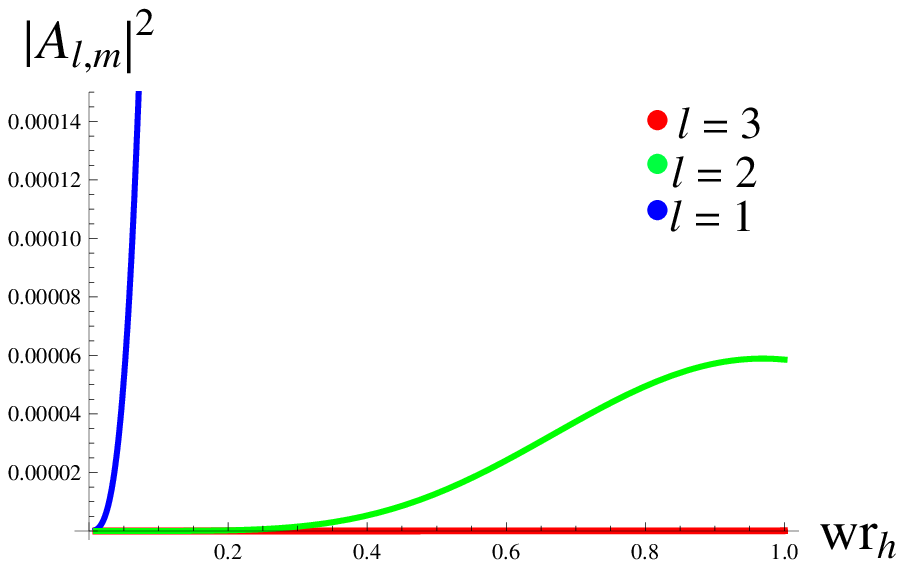}\\
 \caption{Greybody factor for massless scalar field corresponding to
$a_{0}=0.1$, $l=3$ (left plot) and $m=1$, $a_{0}=0.5$ (right plot)
with $r_{0}=0.5$.}\label{3}
\end{figure}
\begin{figure}
  \includegraphics[width=3 in, height=2.5 in]{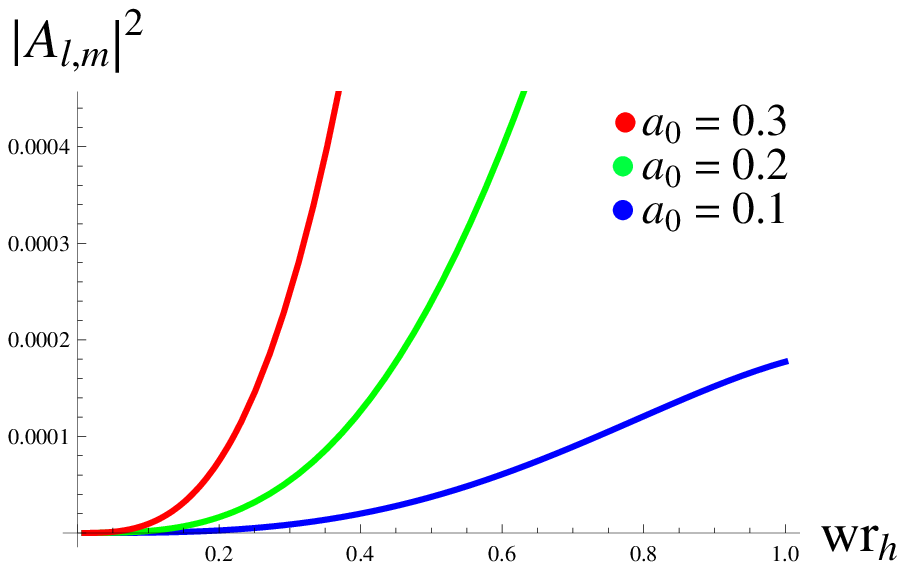}
    \includegraphics[width=3 in, height=2.5 in]{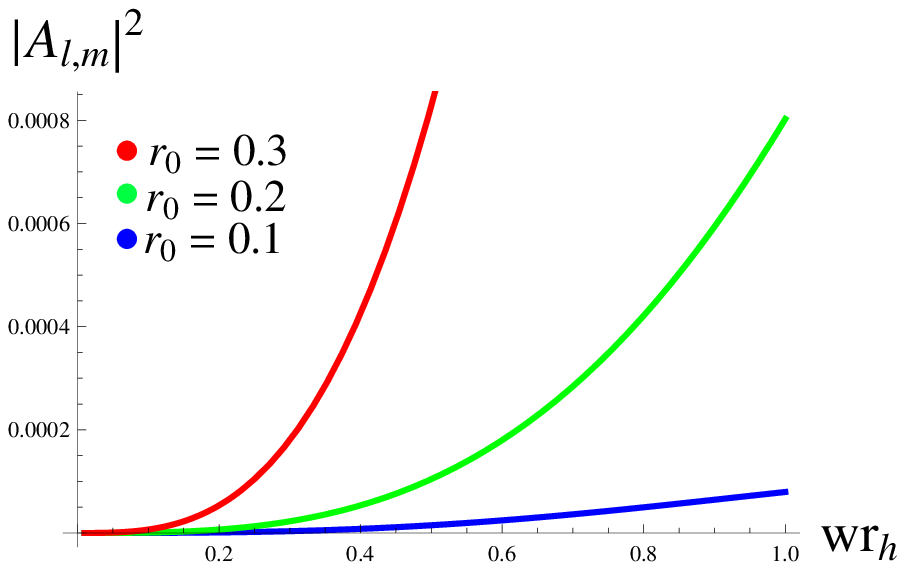}\\
 \caption{Greybody factor for massless scalar field corresponding to
$r_{0}=0.5$ (left plot) and  $a_{0}=0.5$ (right plot) with
$m=l=1$.}\label{4}
\end{figure}

The particle flux, i.e., the total amount of massless scalar
particles discharged per unit time and frequency from a BH is given
by
\begin{eqnarray}\label{49}
\frac{d^2\tilde{N}}{dtdw}=\frac{1}{2\pi}\sum_{l,m}\frac{1}{e^{\frac{k}{T_{H}}-1}}
|\tilde{\emph{A}}_{l,m}|^{2}, \quad
k=w-\frac{ma_{0}}{r_{h}^2+a_{0}^2},
\end{eqnarray}
and Hawking temperature is defined as
\begin{eqnarray}
T_{H}&=&\frac{(1-a_{*}^{2})}{4\pi(1+a_{*}^{2})r_{h}}.
\end{eqnarray}
Moreover, the energy emission rate is found to be
\begin{eqnarray}\label{50}
\frac{d^2\tilde{E}}{dtdw}=\frac{1}{2\pi}\sum_{l,m}\frac{w}{e^{\frac{k}{T_{H}}-1}}
|\tilde{\emph{A}}_{l,m}|^{2},
\end{eqnarray}
which, through Eq.(\ref{48}), gives rise to
\begin{eqnarray}\label{51}
\frac{d^2\tilde{E}}{dtdw}=\frac{1}{2\pi}\sum_{l,m}\frac{w}{e^{\frac{k}{T_{H}}-1}}
\frac{2\iota(\tilde{B}^{*}-\tilde{B})}{\tilde{B}\tilde{B}^{*}
+\iota(\tilde{B}^{*}-\tilde{B})+1}.
\end{eqnarray}
Similarly, we can write differential equation for the angular
momentum emission rate. As the greybody factor depends upon both
particle as well as spacetime properties, therefore it changes the
various emission rates, accordingly. The absorption cross-section
for each partial wave has the form
\begin{eqnarray}\label{52}
\sigma=\frac{\pi}{w^2}\sum_{l,m}|\tilde{\emph{A}}_{l,m}|^{2}.
\end{eqnarray}
Using Eq.(\ref{48}), it follows that
\begin{eqnarray}\label{53}
\sigma=\frac{\pi}{w^2}\sum_{l,m}\frac{2\iota(\tilde{B}^{*}-\tilde{B})}{\tilde{B}\tilde{B}^{*}
+\iota(\tilde{B}^{*}-\tilde{B})+1}.
\end{eqnarray}

\section{Conclusions}

This work is devoted to formulating an analytical form of the
greybody factor for rotating Bardeen regular BH. For this purpose,
we have studied the profile of gravitational barrier which is the
basic reason to generate the absorption probability. We have
evaluated two asymptotic solutions from the radial equation of
motion at different horizons which are valid in low rotation and low
energy region. To obtain a general expression of the greybody
factor, we have stretched these solutions and matched them smoothly
to an intermediate regime. Finally, we have calculated the emission
rates and absorption cross-section for the massless scalar field.
The results are summarized as follows.

We have examined the behavior of angular momentum numbers $(l,m)$
and topological parameters $(a_{0}, r_{0})$ of spacetime on the
effective potential as well as greybody factor. It is found that the
height of gravitational barrier for the massless scalar field
decreases with increasing $a_{0}$ while the larger modes of length
parameter correspond to higher values of the potential barrier which
ultimately reduce the emission rate of Hawking radiation (Figures
\textbf{1-2}). From the graphical display of absorption probability
against the parameter $wr_{h}$, we observe that it attains positive
values in the considered domain and an increase in the parameter $m$
causes an enhancement in the greybody factor (Figure \textbf{3}).
For the orbital angular momentum, partial wave with smaller value
$(l=1)$ dominates in the low approximation limit while all higher
values significantly reduce the greybody factor as in the case of
rotating BH \cite{40}. Moreover, higher modes of rotation and length
parameter yield higher values of the absorption probability (Figure
\textbf{4}). We conclude that the rotation parameter effectively
increases the scalar emission rate for rotating Bardeen BH which is
consistent with the literature \cite{40}.

The greybody factor of a BH can be used to measure its evaporation
rate because it is just the probability of a wave to transmit
through the potential barrier to reach infinity. Based upon our
analysis, it can be seen that regular as well as rotating BH
evaporates more quickly as compared to other BH spacetimes. These
BHs radiate more thermal flux of quantum particles and hence can be
expected to lose their mass, comparatively, in a short span. Hence
the rotating Bardeen BH being the larger emitter of scalar field
particles would shrink and dissipate faster.

According to the no-hair theorem, all astrophysical BH candidates
have correspondence with Kerr BHs, but the existing nature of these
objects still need to be verified \cite{46}. The impact of the
parameter $r_{0}$ on the effective potential, greybody factor and
absorption cross-section presents a good theoretical opportunity to
individualize the rotating Bardeen BH from the Kerr BH and to test
whether astrophysical BH candidates are the actual BHs as predicted
by general relativity.

\vspace{0.5cm}

\textbf{Acknowledgement}

\vspace{0.5cm}

One of us (QM) would like to thank the Higher Education Commission,
Islamabad, Pakistan for its financial support through the
\emph{Indigenous Ph.D. Fellowship, Phase-II, Batch-III}.

\end{document}